\documentclass{www2009-accepted}

\usepackage[utf8]{inputenc}
\usepackage{graphicx}
\usepackage{grffile}
\usepackage[pdftex]{hyperref}
\usepackage[all]{xy}
\usepackage{subfigure}

\begin{document}

\title{The Slashdot Zoo:  Mining \\ a Social Network with Negative Edges}

\numberofauthors{3} 
\author{
\alignauthor
Jérôme Kunegis\\
       \affaddr{DAI-Labor}\\
       \affaddr{Technische Universität Berlin}\\
       \affaddr{Ernst-Reuter-Platz 7}\\
       \affaddr{10587 Berlin, Germany}\\
       \email{kunegis@dai-lab.de}
\alignauthor
Andreas Lommatzsch\\
       \affaddr{DAI-Labor}\\
       \affaddr{Technische Universität Berlin}\\
       \affaddr{Ernst-Reuter-Platz 7}\\
       \affaddr{10587 Berlin, Germany}\\
       \email{andreas@dai-lab.de}
\alignauthor 
Christian Bauckhage\\
       \affaddr{Deutsche Telekom Laboratories}\\
       \affaddr{Ernst-Reuter-Platz 7}\\
       \affaddr{10587 Berlin, Germany}\\
       \email{christian.bauckhage\\@telekom.de}
}

\maketitle
\begin{abstract}
We analyse the corpus of user relationships of the Slashdot
technology news site.
The data was collected from the Slashdot Zoo feature
where users of the website can tag other users as friends and foes,
providing positive and negative endorsements. 
We adapt social network analysis techniques to the problem of negative edge
weights.  
In particular, we consider signed variants 
of global network characteristics such as the clustering coefficient, 
node-level characteristics such as centrality and popularity measures, 
and link-level characteristics such as distances and similarity
measures.
We evaluate these measures on the task of identifying unpopular users,
as well as on the task of predicting the sign of links and show that the
network exhibits multiplicative transitivity which allows algebraic
methods based on matrix multiplication to be used. 
We compare our methods to traditional methods which are only suitable
for positively weighted edges.
\end{abstract}




\section{Introduction}
Social network analysis studies \emph{social networks} by means of
analysing structural relationships between people.
Accordingly, 
social networks are usually modeled using directed graphs, were an edge
between two nodes represents a relationship between two individuals.  
While most social network modeling approaches allow for weighted
edges, the weights are usually restricted to
positive values.  However, some relationships such as distrust and dislike are
inherently negative.  In such cases, the social network contains
\emph{negative edge weights}. 

\begin{figure}[h!]
  \begin{center}
  \includegraphics[width=0.48\textwidth]{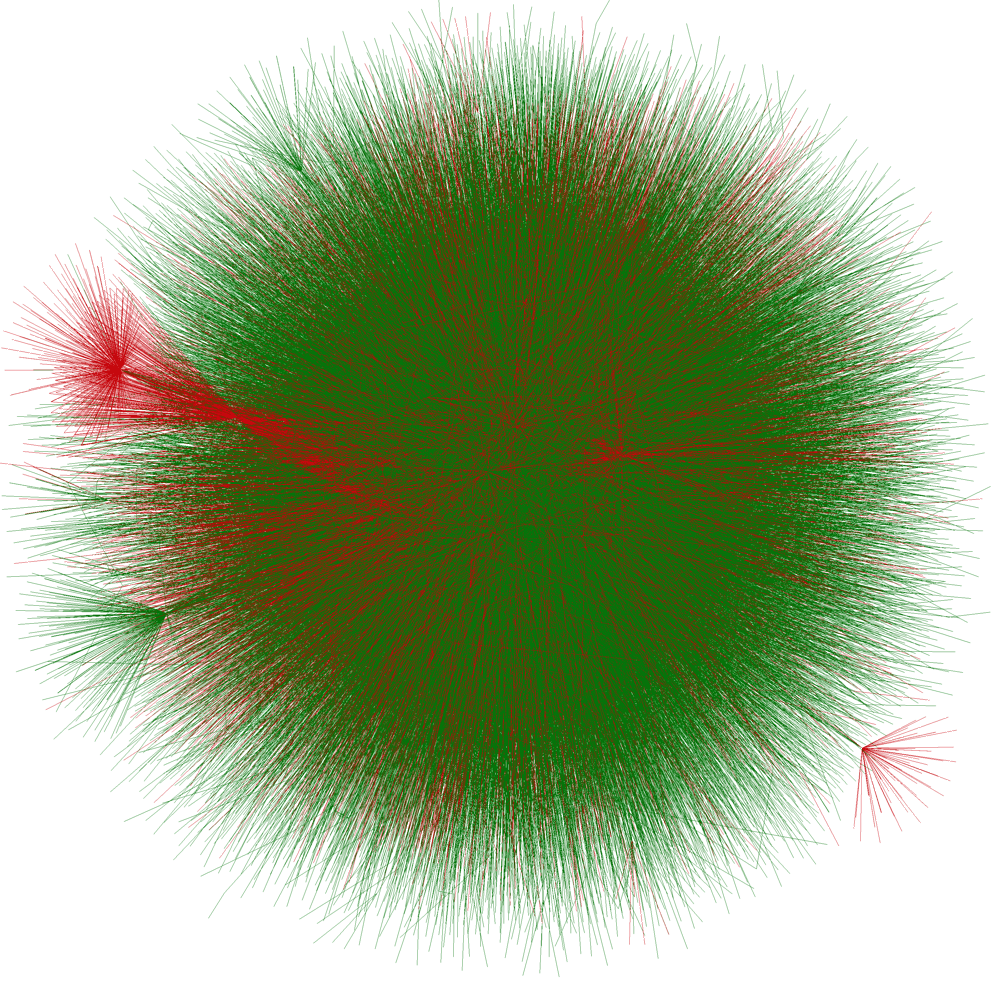}
  \end{center}
  \caption{The Slashdot Zoo corpus represented as a graph, where nodes represent users 
    and edges indicate relationships.  
    The network contains 77,985 nodes and 510,157 edges.
    ``friend'' relationships are
    shown in green edges and ``foe'' relationships in red;
    the orientation of edges is not shown.  The graph is centered at user
  \emph{CmdrTaco}, founder and editor of the site. 
  }
  \label{fig:layout}
\end{figure}

Slashdot\footnote{\url{http://slashdot.org/}} is a technology
news website founded in 1997.
It publishes stories written by editors or submitted by users and allows users to
comment on them. 
In 2002, the site added the \emph{Zoo} feature, which lets
users tag other users as \emph{friends} and \emph{foes}.  
It is therefore an early example of an online service with a social
networking component. 
In constrast to most popular social networking services, 
Slashdot is one of the few sites
that also allows users to rate other users negatively.

In this paper, we systematically study signed versions of methods for
network analysis.  We study the Slashdot Zoo corpus on a
global level, on the level of individual nodes, and on the level of
individual edges. 
On the global level, we analyse measures that characterize the network
as a whole.
We consider the clustering coefficient of the Slashdot Zoo corpus and
propose the signed clustering coefficient and the relative signed
clustering coefficient. 
On the node level, we study measures applying to individual nodes of the
network.  We review signed popularity and centrality measures and evaluate them on
the task of indentifying \emph{troll users}. 
On the edge level, we analyse similarity measures that apply to pairs of
nodes and
evaluate their use for the task of predicting signed links.

The network analysis methods we will present are all based on the
concept of transitivity which stipulates that relations between any two
nodes in the network can be described by paths between the two
nodes.  
In networks with negative edge weights the concept of transitivity has
to take into account negative values.  In the simplest case we can ask:
If there is a path of signed edges between two nodes in the network, what
relation can we induce between the two nodes?  We will show that the
solution to this question is a multiplication rule exemplified by
the phrase \emph{the enemy of my enemy is my friend}.  
By analysing the signed graph on various levels, we will show that this
multiplicative transitivity rule is indeed valid for the Slashdot Zoo.

The rest of the paper is structured as follows.  In
Section~\ref{sec:rel}, we discuss related work.  
Section~\ref{sec:zoo} presents the Slashdot Zoo corpus.  
For network analysis on the global level, Section~\ref{sec:clusco} presents the signed 
clustering coefficient. 
Section~\ref{sec:popularity} discusses various popularity measures on the node level
and evaluates them for the task of indentifying unpopular users. 
Section~\ref{sec:link} reviews distance and
similarity measures on the link level and 
evaluates them on the task of link sign prediction.  
We conclude in Section~\ref{sec:conclusion}. 

\section{Related Work}
\label{sec:rel}
Social network analysis (SNA) has a background in
sociology~\cite{b268}.  
The proliferation of Web~2.0 sites which focus on user participation for content creation resulted
in very large datasets that call for advanced data
mining techniques.
A general discussion of Slashdot can be found in~\cite{b216}. 
The Slashdot discussion threads are studied in~\cite{b331}. 

Most of the web-based social network analysis considers the case of
\emph{unsigned} networks, where edges are either unweighted, or only
weighted with positive values~\cite{b268}.  
Recent studies~\cite{b270}
describe the social network extracted from Essembly, an ideological
discussion site that allows users to mark other users as \emph{friends},
\emph{allies} and \emph{nemeses}, discussing the semantics of the three
relation types.  These works model the different types of edges by means
of three different graphs.  In this paper, we avoid such overhead and
analyse all edges in a single graph with weighted edges. 

Other recent work~\cite{b233} considers the task of
discovering communities from social networks with negative edges.  
However, the negative
edges are only used to separate communities (clusters), and do not serve as
a measure of popularity or similarity.
Signed graphs have been used in anthropology to model friendship and
enmity~\cite{b323}. 

Work on trust networks is by definition concerned with negative edges.  Work
in that field has mostly focused on defining global trust measures
using path lengths or adapting PageRank~\cite{b235,b325,b236}.  In this paper
we compare these approaches to our techniques and find them to yield
inferior results. 
Collaborative filtering aims at predicting or recommending links in a
bipartite user-item graph~\cite{b10}.  The edge weights in such a graph often
admit negative values, indicating a dislike of the item in question.
However, the methods of collaborative filtering cannot immediately be
applied to
social network analysis, because links in the bipartite graph
are not directed and relations between users extracted from a bipartite 
graph are necessarily symmetric. 

The clustering coefficient was first described in~\cite{b228} and
extended to positively weighted edges in~\cite{b269}. 
The task of link prediction in social networks is described
in~\cite{b256} for the case of positive edges.
Distance and similarity in unsigned social networks are described
in~\cite{b262}.  These measures are based on shortest-path distances and
spectral measures such as PageRank and HITS.
Graph kernels are described in~\cite{b191,b263,b156}.
Their application to link prediction and recommendation is covered
in~\cite{b137}.  In all these works however, only positively weighted
edges are considered. 

Balance in a signed graph is defined in~\cite{b355}. 

Variants of Laplacian graph kernels that apply to networks with negative
edges are described in~\cite{kunegis2008a,b351}, where they are
used in an undirected setting for collaborative filtering. 
Centrality and trust measures based on the graph Laplacians are
described in~\cite{b192} for graphs with only positive edge weights.  

\section{The Slashdot Zoo}
\label{sec:zoo}
The Slashdot Zoo corpus we consider in this paper
contains 77,985 users and 510,157 links.
Each link consists of a user who gave an endorsement and a user
who received the endorsement.  
Endorsements can be either positive (``friend'') or negative (``foe'').  
Apart from this distinction, no other
information is available;
in particular, the creation date of endorsements is not known. 

\begin{figure}[b!]
  \centering
  \includegraphics[width=0.39\textwidth]{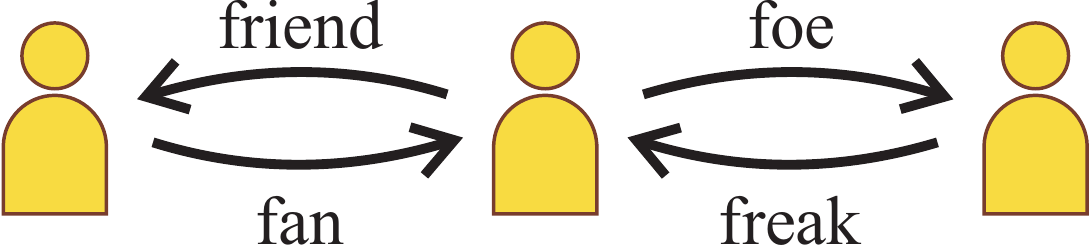}
  \caption{The two types of links allowed in the Slashdot Zoo (friend
  and foe) give rise
  to four kinds of relationships:  friends, fans, foes and freaks.  A
  user is the fan of his friends and the freak of his foes.
  }
  \label{fig:ffff}
\end{figure}

In addition to the terms ``friend'' and ``foe'', Slashdot also uses the
terms ``fan'' and ``freak'':  A user is always the fan of his friends
and the freak of his foes.  Figure~\ref{fig:ffff} summarizes these
relationships. 

Figure~\ref{fig:layout} is a graphical representation of the Slashdot
Zoo corpus.  The sign of an edge is represented by its color, with green
representing the ``friend'' relationship and red representing the
``foe'' relationship.  The graph is centered at user \emph{CmdrTaco}, founder
of Slashdot and active editor. 

Since this data was retrieved between May and October 2008,
our corpus does not represent a true ``snapshot'' of the
network, and may exhibit anomalies.  For instance, since it was not
possible to determine when a user added a tag, some users in our corpus may have more
than 400 friends or foes, although
Slashdot generally limits the number of friends and foes to 200 users
and to 400 users for subscribers. 

Slashdot is known for both having very popular and prominent users on
the one hand, and rather unpopular users on the other hand.
Prominent and popular users of Slashdot include CmdrTaco (Rob Malda,
the founder of Slashdot and a popular editor), John Carmack (prominent
computer game programmer), Bruce Perens (prominent computer programmer
and open source advocate) and CleverNickName (Wil Wheaton, Star Trek
actor).  
In addition, Slashdot is well known for having a rich tradition of
\emph{trolling}, i.e.~the posting of disruptive, false or offensive
information to fool 
and provoke readers.  
The high number of such trolls may explain why the ``foe'' feature is
useful on Slashdot.  It allows for tagging known trolls and reducing
their visibility.   

\subsection{Definitions}
For our graph-based representation of this corpus, we use the
following definitions:
\begin{itemize}
  \item $n$, the number of users 
  \item $u, v$ are specific users
  \item $A \in \{-1, 0, +1\}^{n \times n}$ is the adjacency
    matrix with values
    $A_{uv} = +1$ when user $u$ marked user $v$ as a friend and
    $A_{uv} = -1$ when user $u$ marked user $v$ as a foe.  $A$~is
    sparse, square and asymmetric. 
  \item $\bar A$, the absolute adjacency matrix defined by $\bar A_{ij} =
  |A_{ij}|$ 
  \item $B = A+A^T$, the symmetric adjacency matrix 
  \item $\bar B = \bar A + \bar A^T$, the absolute symmetric adjacency
  matrix
  \item $\bar D$, the absolute diagonal
  degree matrix defined by $\bar D_{ii} = \sum_j |A_{ij}|$
  \item $\bar E$, the absolute symmetric
  diagonal degree matrix defined by $\bar E_{ii} = \sum_j |B_{ij}|$
\end{itemize}

\subsection{Statistics}
\begin{table}
  \centering
  \caption{Statistics about the Slashdot Zoo corpus.  The mean
  friend count and mean fan count are necessarily equal, as do the mean
  foe and freak counts.
  }
  \label{tab:stat}
  \begin{tabular}{|l||r|}
    \hline
    Users & 77,985 \\
    \hline
    Links & 510,157 \\
    Friend links & 388,190 \\
    Foe links & 121,967 \\
    \hline
    Sparsity & 0.000083884 \\
    \hline
    Mean link count & 6.542 \\
    Mean friend/fan count & 4.978 \\
    Mean foe/freak count &  1.564 \\
    \hline
    Median links & 3 \\
    Median friend count & 1 \\
    Median foe count & 0 \\
    Median fans count & 1 \\
    Median freaks count & 1 \\
    \hline
  \end{tabular}
\end{table}

\begin{figure}[t]
  \centering
  \includegraphics[width=0.39\textwidth]{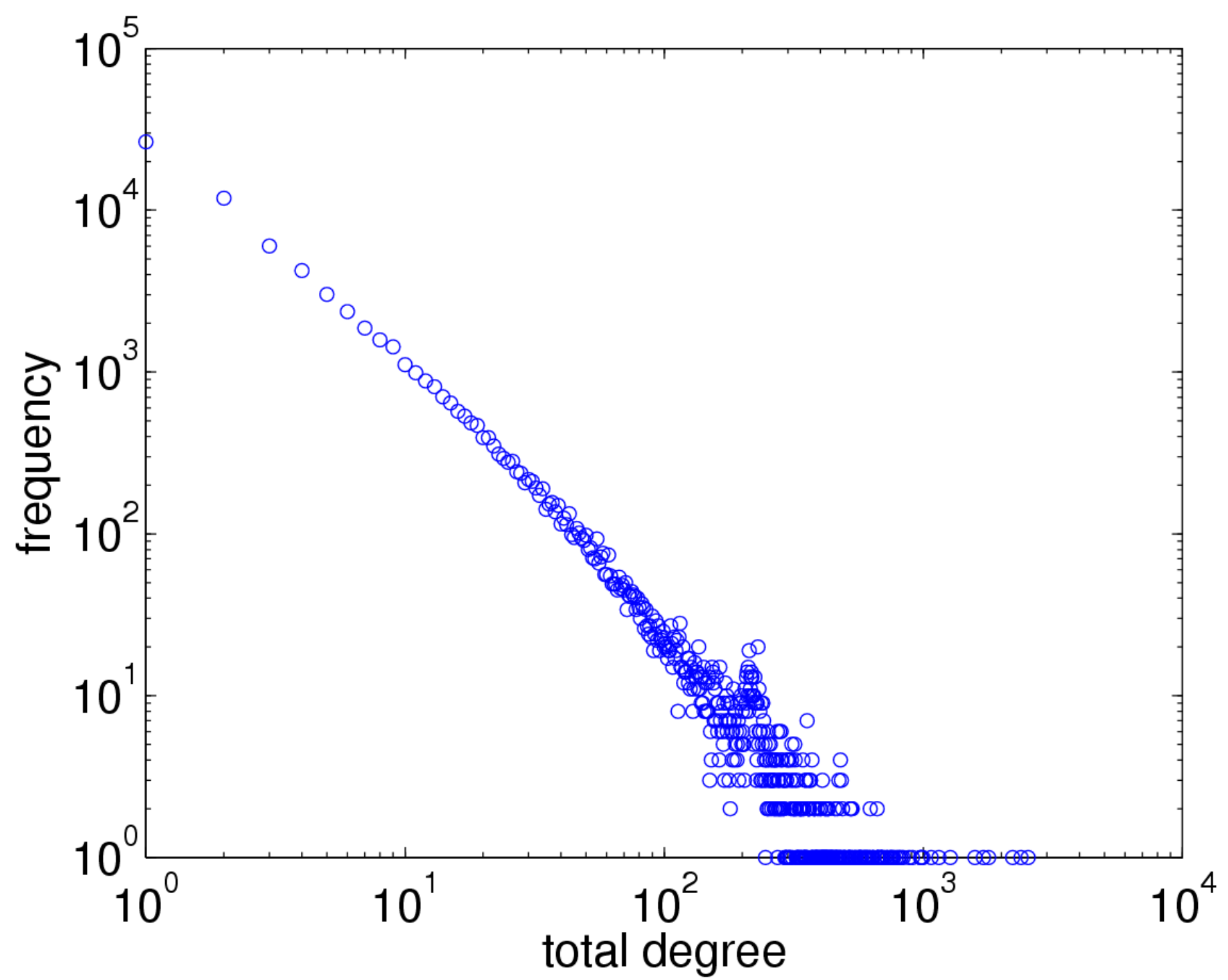}
  \caption{
    Logarithmic plot of the degree distribution showing that the degree
    distribution in the Slashdot Zoo follows a power law. 
    The limit of 200 friends and foes is visible. 
  }
  \label{fig:deg}
\end{figure}

\begin{figure}[t]
  \centering
  \includegraphics[width=0.39\textwidth]{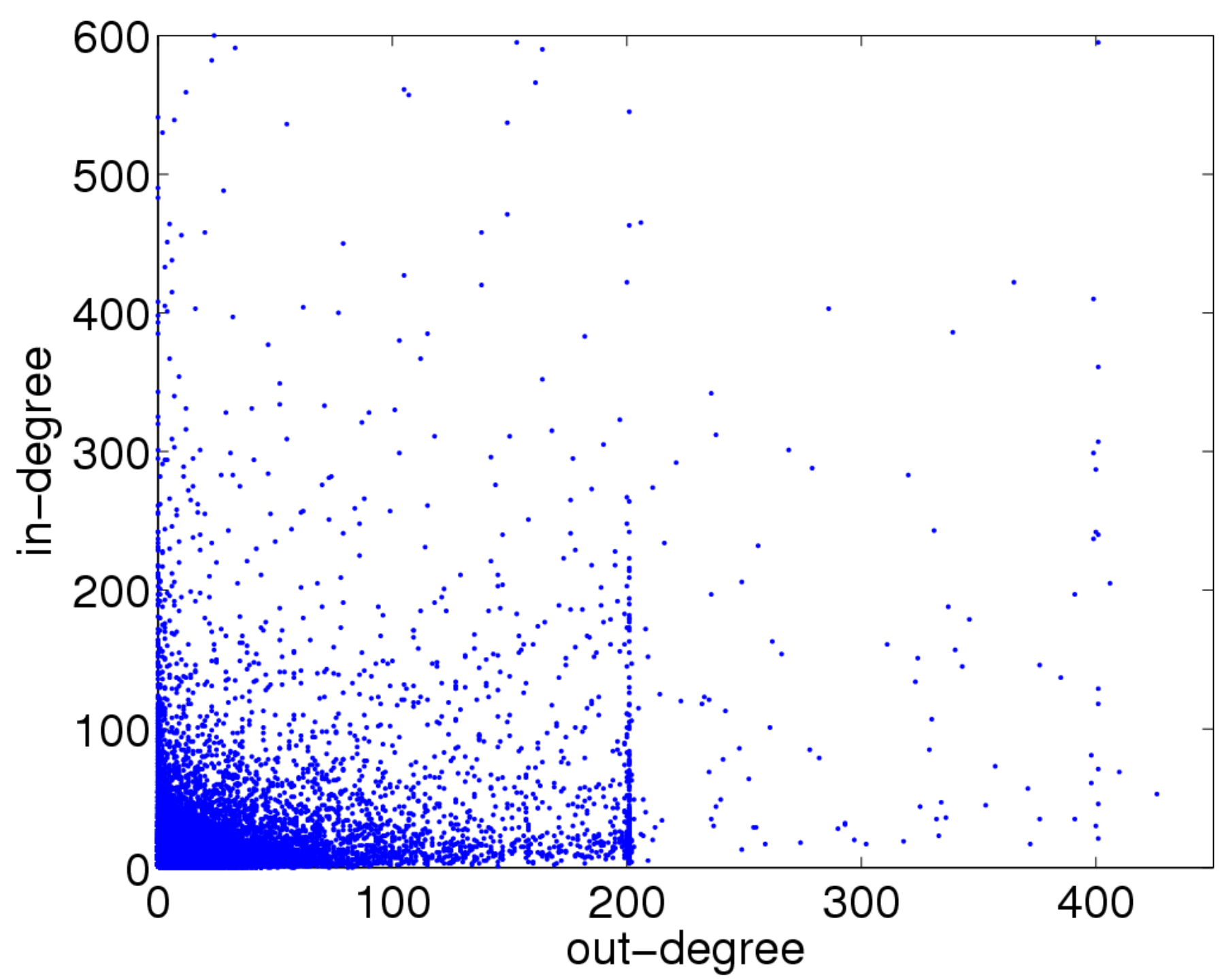}
  \caption{The in-degree plotted against the out-degree.   The limits of 200 and
    400 friends and foes are visible. 
  }
  \label{fig:inout}
\end{figure}
Table~\ref{tab:stat} displays basic statistics of the corpus.
Figure~\ref{fig:deg} shows the degree distributions in the Slashdot
Zoo.  As expected, the degree distribution in the Slashdot Zoo follows a
power law. 
Figure~\ref{fig:inout} shows the in-degree plotted against the
out-degree. 

\begin{table}
  \centering
  \caption{The Slashdot Zoo's graph diameter, radius and mean
    shortest-path distance.  The sign and direction of edges was ignored
    in the calculation of these values.
    In parentheses, we show the average distance in a random graph, as
    defined by Watts and Strogatz.  
  }
  \begin{tabular}{|l||l|}
    \hline
    Diameter & 6 \\
    \hline
    Radius  & 3 \\
    \hline
    Average distance & 3.86 (5.82) \\
    \hline
  \end{tabular}
  \label{tab:net}
\end{table}
Table~\ref{tab:net} shows graph statistics based on shortest path
distances.  All distances were calculated without taking into account
the edge directions and sign.  In parentheses, we show the average
distance as computed in~\cite{b228}.  
As observed in that article, the measured average distance is less than
the average distance in a random graph, confirming that the Slashdot Zoo
is a small-world network. 

\begin{figure}[b!]
  \centering
  \subfigure[PCA by ratings given]{\includegraphics[width=0.23\textwidth]{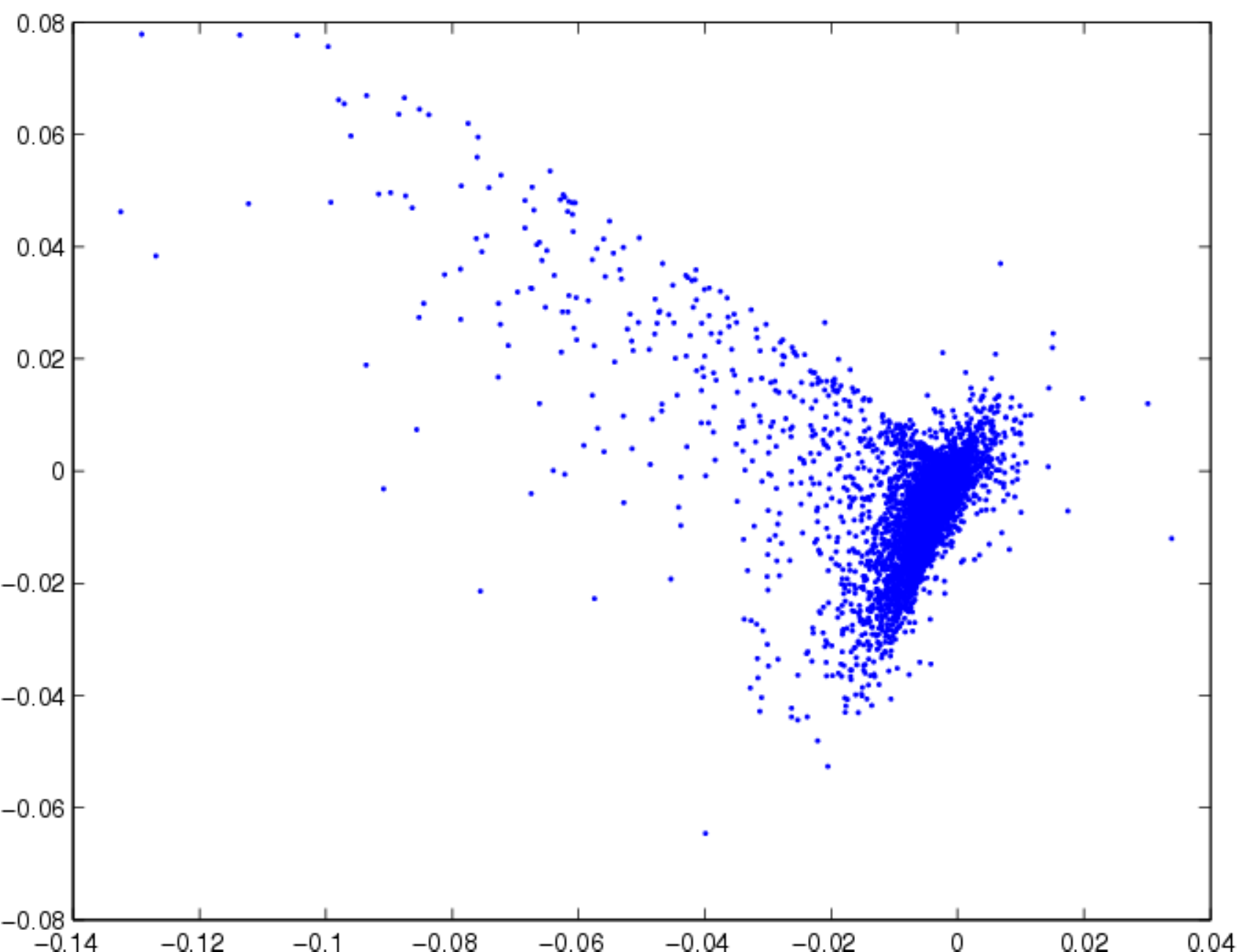}}
  \subfigure[PCA by ratings received]{\includegraphics[width=0.23\textwidth]{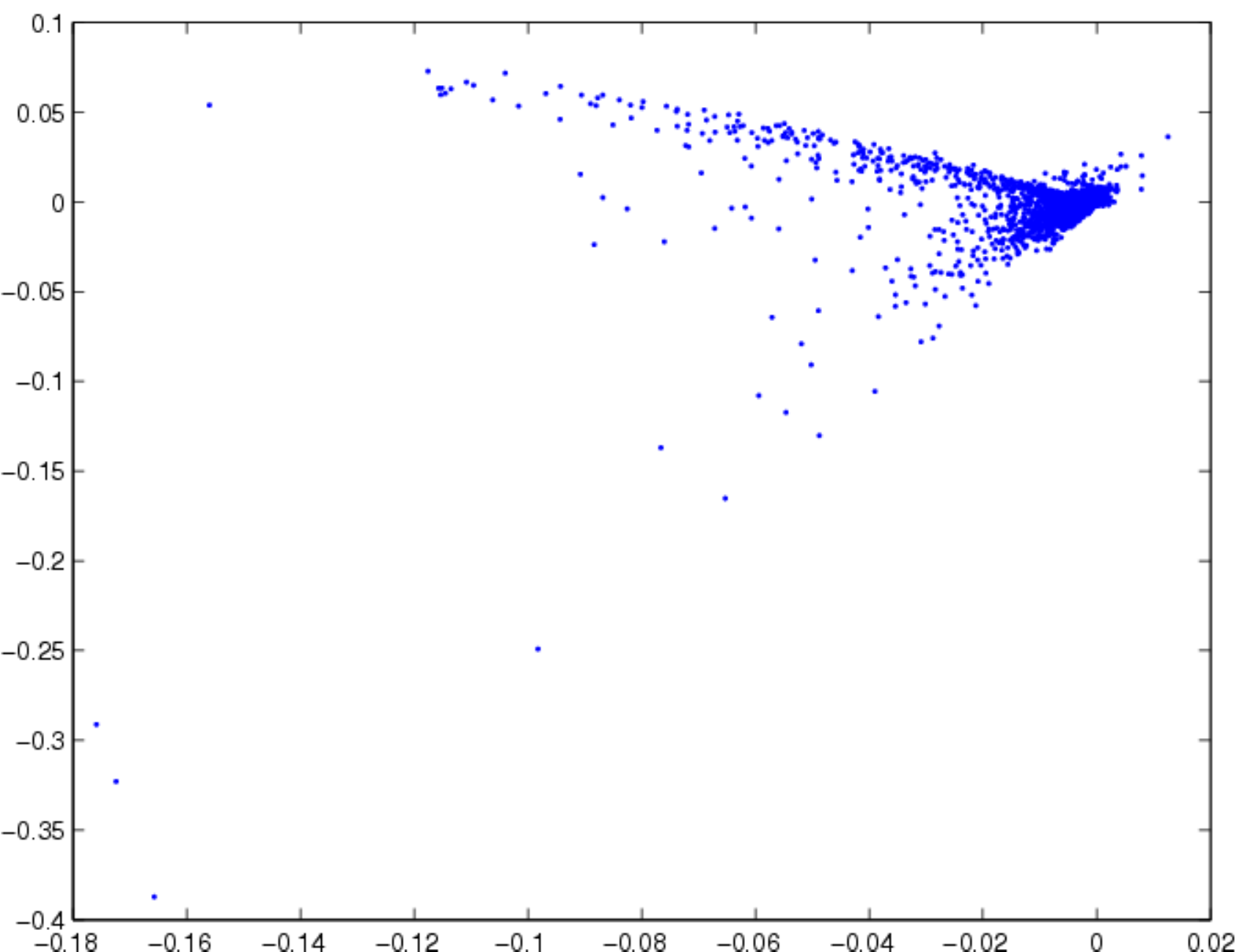}}
  \subfigure[Laplacian PCA]{\includegraphics[width=0.23\textwidth]{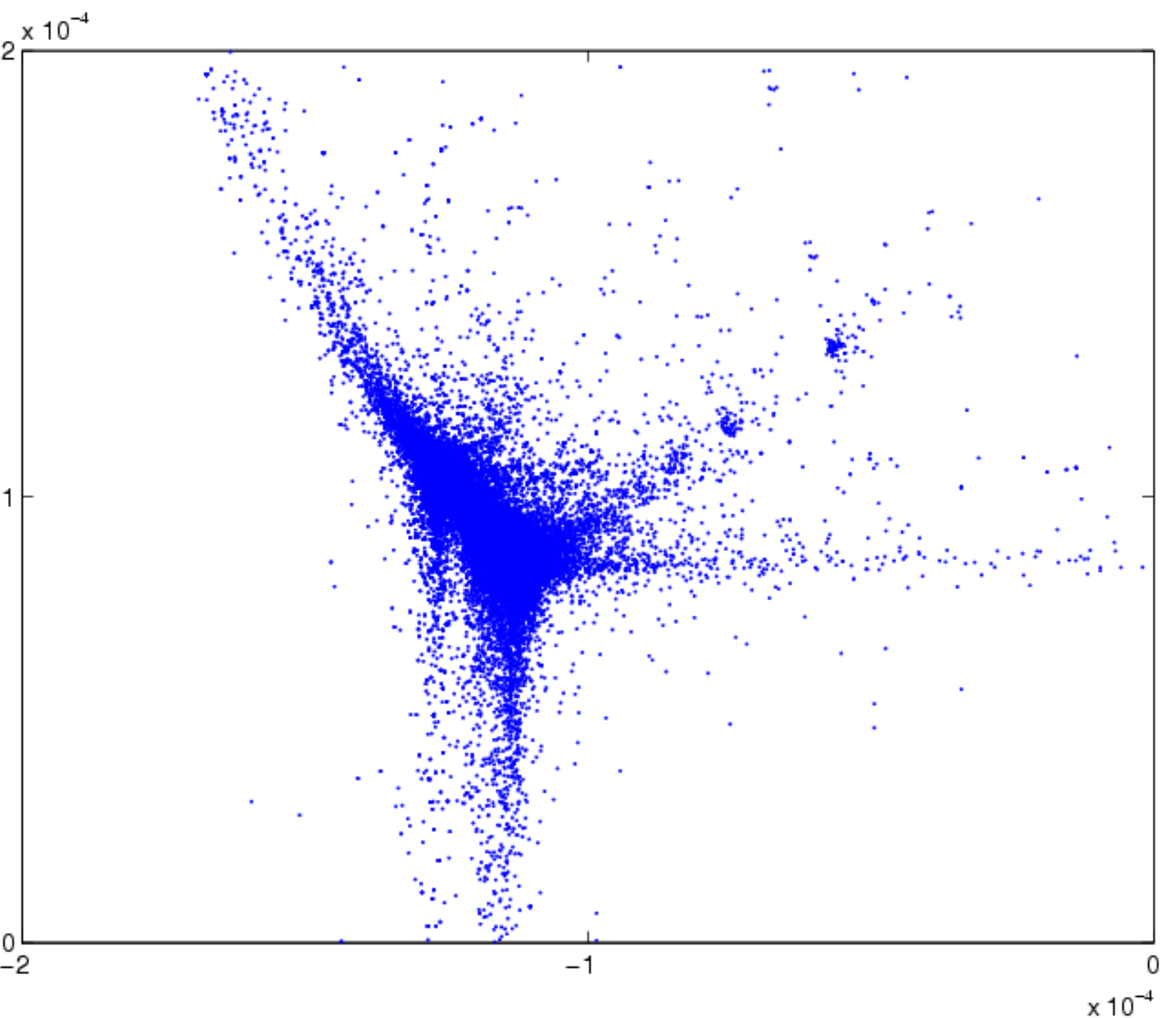}}
  \caption{Three variations on principal component analysis (PCA).  In these
  graphs, every user is represented by a point.  (a) Users are
  represented by the ratings they gave to other users.  (b) Users are
  represented by the ratings they received from other users.  (c)
  Laplacian PCA approximating the resistance distance in the unsigned
  underlying graph.   
  }
  \label{fig:pca}
\end{figure}
Figure~\ref{fig:pca} shows three variants of principal components
analysis, visualizing the dataset.  The first two plots show the users
plotted by the ratings they have given and the ratings they have
received, respectively.  The third plot shows the principal component
analysis as given by the Laplacian kernel of Equation~\ref{eq:lap} in
Section~\ref{sec:link}.  As we will see, the projection using the
Laplacian kernel gives a plot preserving the resistance distance between
nodes in the network. 

In the next three sections, we study the Slashdot Zoo corpus on the
global, node and link levels.

\section{Clustering Coefficient}
\label{sec:clusco}
To test our hypothesis of multiplicative transitivity, we begin by
defining and studying a global network statistic that denotes to what
extent the \emph{multiplication rule} is valid in a signed network.
To that end,
we extend the clustering coefficient to networks with
signed, directed edges to give the signed clustering coefficient, 
introduce the relative signed clustering coefficient, and give the
values observed for the Slashdot Zoo corpus.  
As we will see,  the signed clustering coefficient denotes the 
transitivity of edge signs, giving an indication whether our
multiplicative transitivity assumption is justified.  
The two next sections will then study popularity and link prediction
algorithms that directly make use of multiplicative transitivity. 
We begin by defining multiplicative transitivity.

{\sc Definition.}
{\it
A signed network exhibits multiplicative transitivity when any two
incident edges tend to be completed by a third edge having as a weight
the product of the two edges' weights.
}

Multiplicative transitivity is motivated by the fact that triangles of
users connected by an even number of negatively weighted edges can be
considered \emph{balanced}~\cite{b323}, which can be summarized by the
phrase \emph{the enemy of my enemy is my friend} and its permutations. 
As we will see in later sections, the assumption of transitive
multiplicativity lends itself to using algebraic methods based on the
adjacency graph of the network.
To see why this is true, consider that the square $A^2$ of the signed
adjacency matrix contains at its entry $(i,j)$ a sum of paths of length
two between $i$ and $j$ weighted positively or negatively depending on
whether a third positive edge between $i$ and $j$ would lead to a
balanced or unbalanced triangle. 
As a measure of multiplicative transitivity, this section begins by proposing the
signed clustering coefficient.

The clustering coefficient was introduced in~\cite{b228}.
An extension was
proposed in~\cite{b269} that works with positively weighted edges. 
The clustering coefficient is a characteristic number of a graph taking
values between zero and one, denoting the tendency of the graph nodes
to form small clusters.  
The signed clustering coefficient we 
define denotes the tendency of small clusters to be \emph{coherent}.
Therefore, the signed clustering coefficient will take on values between
$-1$ and $+1$.  The relative signed clustering coefficient will be
defined as the quotient between the two. 

\begin{figure}[!b]
  \centerline{\xymatrix @R-31pt @C-2pt {
     & {\bullet} \ar@{-}[rr]^{c} \ar@{-}[rdd]_{a} & & {\bullet} &      & {\bullet} \ar@{->}[rr]^{c} \ar@{->}[rdd]_{a} & & {\bullet} \\ 
    {a)} & & & &   {b)}  \\
    & & {\bullet} \ar@{-}[ruu]_{b} & &     & & {\bullet} \ar@{->}[ruu]_{b} & 
    }}
  \vspace{0.3cm}
  \centerline{\xymatrix @R-31pt @C-2pt {
     & {\bullet} \ar@{-}[rr]^{c = ab} \ar@{-}[rdd]_{a} & & {\bullet} &      & {\bullet} \ar@{->}[rr]^{c = ab} \ar@{->}[rdd]_{a} & & {\bullet} \\
    {c)} & & & &    {d)} \\
    & & {\bullet} \ar@{-}[ruu]_{b} & &     & & {\bullet} \ar@{->}[ruu]_{b} & 
    }}
  \caption{The four kinds of clustering coefficients.  a) Regular
  clustering coefficient.  b) Directed clustering coefficient. c) Signed
  clustering coefficient.  d) Signed directed clustering coefficient.
  Edge $c$ is counted when edges $a$ and $b$ are present,
  and for the signed variants, weighted by $\mathrm{sgn}(abc)$.}
  \label{fig:clusco}
\end{figure}
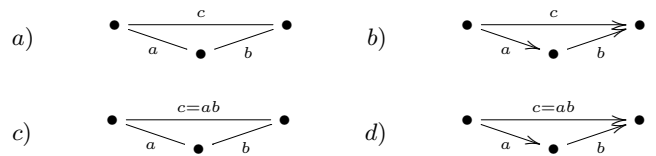
The clustering coefficient is defined as the proportion of all incident
edge pairs that are completed by a third edge to form a triangle.
Figure~\ref{fig:clusco} gives an illustration. 
Alternatively, the clustering coefficient can be defined algebraically
using the adjacency matrix of the network.  
Given an undirected, unweighted loopless graph $G$ with unsigned
adjacency matrix $\bar A$, its clustering coefficient is given by
\begin{eqnarray}
  C(G) &=& \frac{\parallel \bar A \circ \bar A^2\parallel _+}{\parallel
  \bar A^2\parallel _+} \label{eq:cc}
\end{eqnarray}
where $A \circ B$ represents the Hadamard product (entrywise product) of
two matrices, and $\parallel A\parallel _+$ denotes the sum of all
matrix elements. 
Since $G$ is undirected and unweighted, $\bar A$ is a nonnegative symmetric matrix
with a diagonal of zero.

If we now suppose that $G$ is a directed
graph, $\bar A$ will no longer be symmetric, but Expression~\ref{eq:cc} is
still defined.  In this case, only edges completing two directed edges
in the correct orientation will be counted.

To extend the clustering coefficient to negative edges, we assume a
multiplication rule for two incident signed edges.
The multiplication rule captures the intuition that \emph{the enemy of my
enemy is my friend}. 
As shown in Figure~\ref{fig:clusco}, an edge $c$ completing two incident
edges $a$ and $b$ to form a triangle must fulfill the equation $c=ab$. 
Because $A$ already contains the value $-1$ for negative edges,
the square $A^2$ will contain the sum of such products, leading to 
the following expression for calculating the
signed clustering coefficient:
\begin{eqnarray}
  C_s(G) &=& \frac{\parallel A \circ A^2\parallel _+}{\parallel\bar
  A^2\parallel _+} \label{eq:ccs}
\end{eqnarray}
Therefore, the signed clustering coefficient denotes to what extent
the graph exhibits multiplicative transitivity. 
In actual social networks, we expect it to be positive.

Additionally, we define the relative signed clustering coefficient as the quotient
of the signed and unsigned clustering coefficients.  
Graphs with a high relative signed clustering
coefficient are thus graphs for which the multiplication rule applies.
\begin{eqnarray}
  S(G) &=& \frac{C_s(G)}{C(G)} = \frac{\parallel A \circ A^2\parallel _+}{\parallel \bar A \circ \bar A^2\parallel _+} \label{eq:scc}
\end{eqnarray}
The relative signed clustering coefficient takes on values between $-1$ and $+1$.
It is $+1$ when all triangles are oriented coherently.  In networks with
negative relative signed clustering coefficients, the sign multiplication rule
does not hold. 
The directed signed clustering coefficient and directed relative signed clustering
coefficient can be defined analogously with Expressions~(\ref{eq:ccs})
and~(\ref{eq:scc}).  

The signed clustering coefficient and relative signed clustering coefficient are
zero in random networks, when the sign of edges is distributed equally. 
The signed clustering coefficients are by definition smaller than
their unsigned counterparts.
The unsigned clustering coefficient overestimates the number of
``correctly'' placed edges by ignoring their weight.

\begin{table}
  \centering
  \caption{The values for all variants of the clustering
    coefficient for the Slashdot Zoo corpus.  
    In parentheses, we give the clustering coefficient of
    a random graph of the same size, computed using the method of Watts and
    Strogatz, which applies only to unsigned networks.  
    Together with the results from Table~\ref{tab:net}, these results
    show that the Slashdot Zoo exhibits the small-world phenomenon. 
  }
  \begin{tabular}{|l||l|}
    \hline
    Clustering coefficient & 3.19\% (0.0095\%) \\
    Directed clustering coefficient & 5.62\% (0.0191\%) \\
    Signed clustering coefficient & 2.44\% \\
    Directed signed clustering coefficient & 4.44\% \\
    \hline
   \end{tabular}
  \begin{tabular}{|l||l|}
    \hline
    Relative signed clustering coefficient & 76.4\% \\
    Directed relative signed clustering coefficient & 79.0\% \\
    \hline
  \end{tabular}
  \label{tab:cc}
\end{table}
Table~\ref{tab:cc} gives all four variants of the clustering coefficient
measured in the Slashdot Zoo corpus, along with the relative signed
clustering coefficients.   
We also give the clustering coefficient of a random graph of the same
size, as described in~\cite{b228}.  The clustering coefficient of the
Slashdot Zoo is significantly larger than that of a random graph of
equal size.  Together with the obsersvation in Table~\ref{tab:net}
that the average distance between nodes is less than in that of a random
graph, we follow Watts and Strogatz~\cite{b228} and conclude that the
Slashdot Zoo exhibits the small-world phenomenon. 
The high values for the relative clustering coefficients show that
our multiplication rule is valid in the Slashdot Zoo, and justifies the
use of algebraic methods in the next two sections for popularity and link
prediction methods. 

\section{Popularity and Centrality}
\label{sec:popularity}
In this section, we review measures that apply to single nodes in the
network.  We show how standard centrality and popularity measures can be
extended to the case of negative edges, and how these perform on the
task of identifying unpopular users.

Centrality in the broadest sense is a measure computed for each node in
a graph, denoting to what extent the node is \emph{central} to the
graph.  In social networks, such measures are also called
\emph{importances}.  
Central nodes are usually well connected to other nodes, and at a short
distance to most other nodes.  On the other hand, decentral nodes
are poorly connected and at a greater distance to other nodes.  
Centrality measures can be defined by taking a given distance measure on
the graph, and measuring the average distance from all other nodes.
Other centrality measures are defined by considering a flow or diffusion
process in the network, and computing the amount of flow going through
each node.
The notion of trust is usually personalized, i.e. it applies to pairs of
users.  If however trust is taken as a global measure, then it
corresponds largely to the concept of popularity we study here. 

While centrality can be defined independently of edge signs, giving
measures of who is central regardless of any sign, we will focus on
centrality measures that are signed, and can be used to identify both
top and bottom users, corresponding to well-liked and much-unliked
users.  
We will first describe the various centrality measures we used, then
give the top six users for each, and then evaluate the measures on the
task of identifying trolls.

\subsection{Popularity and Centrality Measures}
We now describe the centrality and popularity measures we evaluated. 

\subsubsection{Fans Minus Freaks (FMF)} 
As a baseline popularity measure, we use the number of fans and freaks a user
has for calculating his reputation.  We subtract the number
of freaks from the number of fans, giving a signed number.  
We expect popular users to have a high number of fans and unpopular
users a
high number of freaks, making this measure a valid indicator of popularity.
While this measure is simple, it can also be exploited easily using
multiple accounts.  Malicious users may create accounts
with the sole purpose of marking oneself as a friend, just as malicious
websites may try to create linkfarms in order to boost their ranking in
search engines.  

\subsubsection{PageRank (PR)}
PageRank is a spectral popularity measure defined on directed graphs with nonnegative
edge weights~\cite{b271}. 
It models the path of a random ``surfer''
following the directed edges of the graph randomly, and ``teleporting''
to a random node at randomly chosen intervals. 
Equivalently, it can be defined as the dominant left eigenvector of the
\emph{Google matrix} $G$, given by
\begin{eqnarray}
  G &=& (1-\alpha) \bar D^{-1} \bar A + (\alpha/n) J_{n \times n}
\end{eqnarray}
where $J_{n \times n}$ is a matrix full of ones of the specified size,
and $0 < \alpha < 1$ is the teleportation parameter.  The matrix $G$ is
left-stochastic (each row sums to one). 
Since PageRank only applies to nonnegative edge weights, we have to use
the unsigned adjacency matrix $\bar A$ instead of $A$. 
The resulting rank is thus not an
indication of popularity, but more an indication of centrality, denoting the
tendency of users to be central, without distinguishing friend and foe
links.  

\subsubsection{Signed Spectral Ranking (SR)}
PageRank was extended to the case of negative edge weights several
times~\cite{b235,b236}, modeling the resulting ranking 
as a measure of popularity.  We compute the signed spectral ranking as the
dominant left eigenvector of the signed matrix $G_s$:
\begin{eqnarray}
  G_s &=& (1-\alpha) \bar D^{-1} A + (\alpha/n) J_{n \times n}
\end{eqnarray}
The resulting popularity measure admits both positive
and negative values, and represents a measure of popularity in the network,
with positive edges corresponding to a positive endorsement and negative
edges to negative endorsements.  This interpretation is consistent with
the semantics of the ``friend'' and ``foe'' relationships.

\subsubsection{Signed Symmetric Spectral Ranking (SSR)}
We apply spectral ranking to the symmetric matrix $B = A + A^T$.  Thus,
outlinks by users are counted towards the popularity of a user.  While this
measure could be exploited by users to gain a high rank, it models the
idea that ``popular'' users will refrain from having many foes, and that
negative edges are more prevalent between unpopular users, regardless of
the edge direction. 

\subsubsection{Negative Rank (NR)}
\begin{figure}
  \centering
    \includegraphics[width=0.44\textwidth]{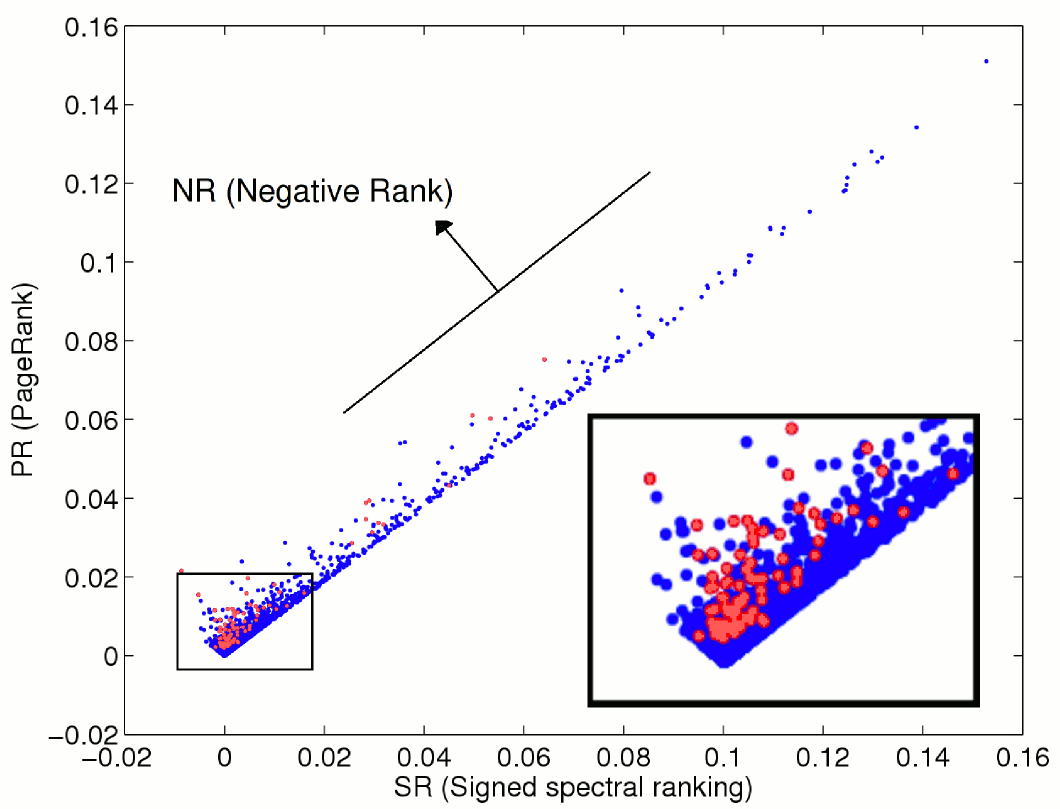}
  \caption{A scatter plot of all users showing PageRank against Signed
    Spectral Rank, with known trolls marked in red.
    While the dominant trend is a positive correlation, we
    observe that the users with lowest Signed Spectral Rank have a
    (relatively) high PageRank.
    For most users, both ranks are almost equal while for trolls, 
    Signed Spectral Rank is smaller. 
  }
  \label{fig:v}
\end{figure}

Comparing PageRank with Signed Spectral Ranking in Figure~\ref{fig:v}
shows that these two rankings correlate highly.  We observe that for
most users, both measures are almost equal.  For known trolls however,
the Signed Spectral Rank seems to be much less than PageRank.
Therefore, we propose to subtract the two to define the Negative Rank:
\begin{eqnarray}
  \mathrm{NR} &=& \mathrm{SR} - \beta \cdot \mathrm{PR}
\end{eqnarray}
The parameter $\beta$ determines the influence of PageRank on the
ranking, and SR and PR are both of unit-length.  
For $\beta = 0$, Negative Rank corresponds to the Signed
Spectral Rank.  We will first assume a value of $\beta = 1$ for now,
and study the influence of $\beta$ later. 

\subsection{Top and Central Users}
The top six users for each popularity and centrality  measure are shown in
Table~\ref{tab:top}.  
We do not show the bottom users to discourage users abusing the system.
However, we note that in the list of users with most freaks, place five
is taken by user JonKatz, an editor that left Slashdot in 2002.  
The bottom users in the Fans Minus Freaks and Signed Spectral Clustering
measures are known Slashdot trolls.  The bottom users for Signed Symmetric Spectral
Rank and PageRank are not known trolls.

The top users for Negative Rank are actually neither prominent nor very
central in the Slashdot community. We explain this by the fact that
many users have a high Negative Rank of approximately zero, just as if
we used a low number of freaks as a measure for finding popular users:
While users without freaks may be popular, there are so many of them
that the metric is not useful to find the very popular users.  However,
as we will see, Negative Rank is very good at identifying unpopular
users. 

\begin{table}[t!]
  \caption{
    The top six users for the five centrality and popularity measures.
    Five of the six top users by the Fans Minus Freaks measure are
    prominent, while the top users by the spectral measures are frequent
    but not prominent users.  We conjecture that these frequent users are more
    ``in the community'' than the prominent users, making the spectral
    centrality and popularity measures more suited for identifying central
    users on web communities.
    While Negative Rank is able to indentify unpopular users in the
    Slashdot community, it is not useful for finding central or
    prominent users. 
  }
  \centering
\begin{tabular}{|rr|}
\hline
\multicolumn{ 2}{|c|}{Fans Minus Freaks} \\
\hline
\hline
CleverNickName &       2460 \\

Bruce Perens &       2143 \\

  CmdrTaco &       2005 \\

John Carmack &       1663 \\

NewYorkCountryLawyer &       1179 \\

\$\$\$\$\$exyGal &       1170 \\
\hline
\hline
\multicolumn{ 2}{|c|}{PageRank} \\
\hline
\hline
  FortKnox &     0.1510 \\

SamTheButcher &     0.1342 \\

Ethelred Unraed &     0.1280 \\

      turg &     0.1266 \\

Some Woman &     0.1254 \\

  gmhowell &     0.1247 \\
\hline
\hline
\multicolumn{ 2}{|c|}{Signed Spectral Ranking} \\
\hline
\hline
  FortKnox &    0.1527 \\

SamTheButcher &    0.1388 \\

      turg &    0.1319 \\

Some Woman &    0.1310 \\

Ethelred Unraed &    0.1297 \\

  gmhowell &    0.1263 \\
\hline
\hline
\multicolumn{ 2}{|c|}{Signed Symmetric Spectral Ranking} \\
\hline
\hline
Ethelred Unraed &    0.1395 \\

  johndiii &    0.1376 \\

  FortKnox &    0.1358 \\

      turg &    0.1351 \\

Some Woman &    0.1309 \\

SamTheButcher &    0.1276 \\
\hline
\hline
\multicolumn{ 2}{|c|}{Negative Rank} \\
\hline
\hline
   johndiii  & 0.006279 \\
   SolemnDragon &  0.006181 \\
   Some Woman &  0.005515 \\
   KshGoddess & 0.005487 \\
   turg &  0.005299 \\
   btlzu2 & 0.005179 \\
\hline
\end{tabular}  
  \label{tab:top}
\end{table}
Table~\ref{tab:top} shows that the top five users for Fans Minus Freaks
are prominent.  For the other measures however, the top users are
frequent users, but not prominent.  We suspect that these users have a
greater involvement in the Slashdot community, leading to a tightly
clustered network of friends.  Prominent users such as \emph{CmdrTaco} on the other hand may
have many fans, but are not central to the Slashdot community. 

\begin{figure}
    \includegraphics[width=0.48\textwidth]{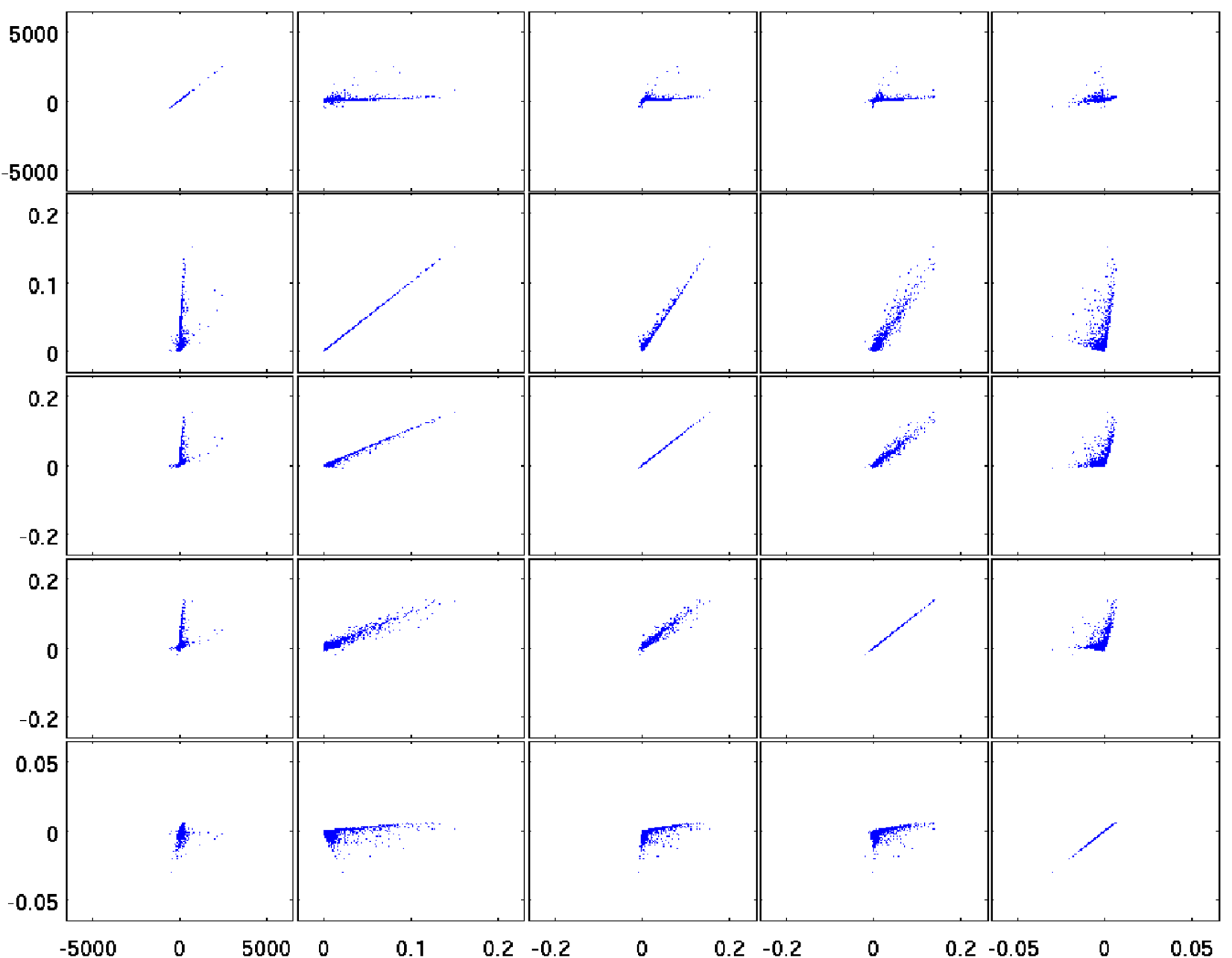}
    \caption{The five popularity and centrality measures plotted against each
      other.  From top/left to bottom/right the figure shows:  Fans Minus Freaks,
      PageRank, 
      Signed Spectral Rank, Signed Symmetric Spectral Rank and
      Negative Rank.  
      A positive correlation can be observed between the spectral
      methods (excluding Negative rank), while Fans Minus Freaks and
      Negative Rank do not correlate significantly with any other measure. 
    }
    \label{fig:plotmatrix}
\end{figure}
Figure~\ref{fig:plotmatrix} plots the five centrality and popularity measures
against each other.  We observe a positive correlation between the three
spectral methods, and no significant correlation between
Fans Minus Freaks, Negative Rank and the other methods.  

\subsection{Predicting Trolls}
To evaluate whether our popularity measures are able to make good predictions
for unpopular users, we evaluate them on the task of identifying
trolls.  Such an automatic identification of trolls could for instance be
used to maintain a list of trolls more accurately than maintaining it by
hand. 

As a benchmark, we use the foes of the user \emph{No More Trolls}.  This
user account was created with the purpose of collecting names of trolls.
Any Slashdot user can
tag this user as his friend, and then, in the settings, give a
malus to foes-of-a-friend, which reduces the visibility of \emph{No More
Trolls}' foes.
We use only known trolls having a minimum of 20 incident
edges, as trolls with fewer edges would be hard to predict.  There are
162 known
trolls found this way. 

The task consists of predicting who these trolls are by using a
popularity measure defined in the corpus excluding edges incident to \emph{No
More Trolls}. 
We use each measure for predicting the unpopular users, and
give the mean average precision (MAP) for each~\cite{b272}. 

In addition to the five popularity measures above, we use the negated number
of freaks as a further popularity measure.  For Negative Rank, we use
$\beta=1$. 

\begin{figure}[t!]
  \centering
  \includegraphics[width=0.35\textwidth]{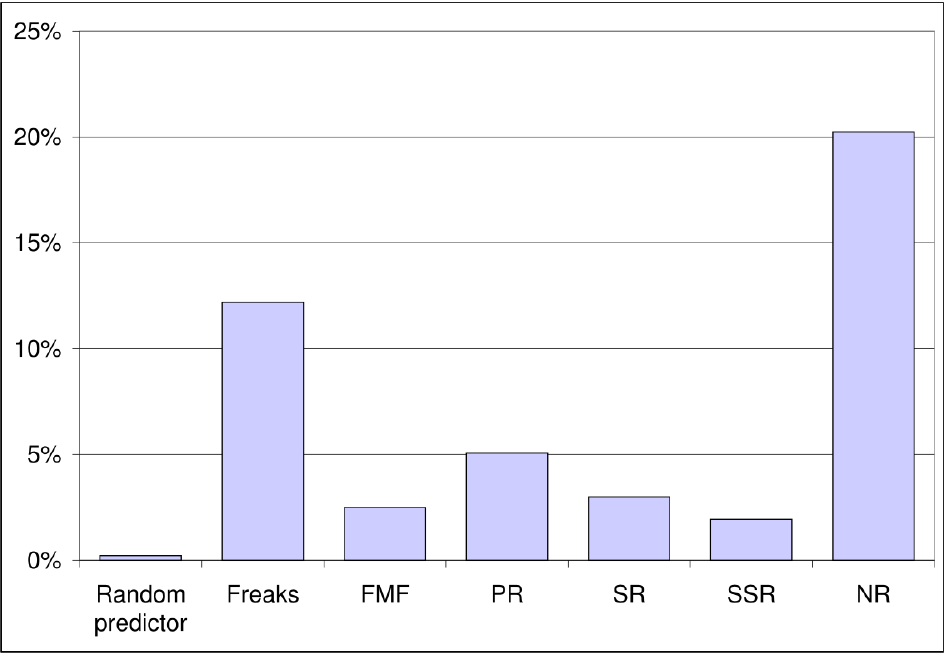}
  \caption{The mean average precision of the four popularity and centrality
    measures at the task of identifying trolls.  
    For comparison, we show
    the precision of predicting trolls at random.  Freaks:  number of
    freaks.  FMF:  Friends Minus Freaks.  PR:  PageRank. SR:  Signed
    Spectral Ranking.  SSR:  Signed Symmetric Ranking.  NR:  Negative
    Rank. 
    Negative Rank performs significantly better that all other popularity
    measures at the task of identifying Slashdot trolls. 
  }
  \label{fig:bar}
\end{figure}
The results are shown in Figure~\ref{fig:bar}.  
We observe Negative Rank to perform best.  We also observe that the
negated number of freaks performs better than any spectral measures
except Negative Rank.  Signed spectral ranking alone performs very
badly, indicating that taken alone, it is not a good indicator of
popularity, but has to be combined with PageRank to give Negative Rank. 

\begin{figure}[t!]
  \centering
  \includegraphics[width=0.28\textwidth]{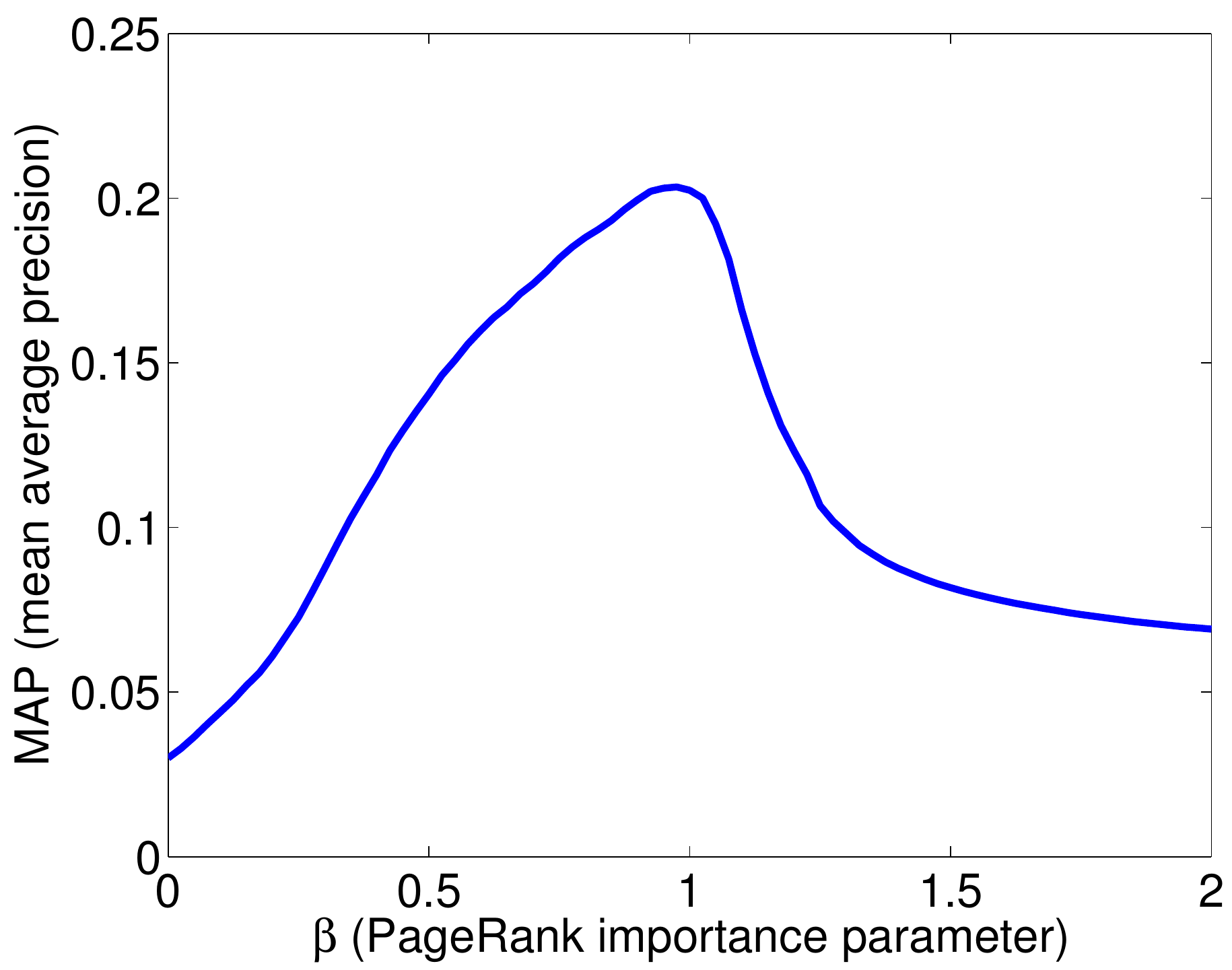}
  \caption{The performance of Negative Rank in function of the parameter
  $\beta$.  We observe that empirically, the best performance is
  attained at $\beta=1$. 
  }
  \label{fig:prst}
\end{figure}
We study the influence of the parameter $\beta$ on the performance of
Negative Rank in Figure~\ref{fig:prst}.  We observe that the optimal
performance is attained at $\beta=1$, confirming our previous
assessment. 

\section{Link Prediction}
\label{sec:link}
In this section, we analyse measures defined on node pairs.  This type
of measure covers similarities, distances, mutual trust, and other
related concepts.  These measures are usually the most important in
social network studies, first because they describe the most structure
in the network, second because they can be applied to link prediction
and recommendation. 
Since most link prediction methods can be formulated algebraically using
matrix multiplication, we will be able to verify our assumption of
multiplicative transitivity. 

We will thus focus on the task of predicting links in signed social
networks. In particular, we cover the prediction of the \emph{sign} of
edges.  Because link prediction algorithms defined for networks with
nonnegative edge weights are only able to predict the presence or
absence of an edge, we will have to define new link prediction
algorithms that take into account the edge sign, and are able to predict
missing edges' sign. 

In social networking websites, link prediction allows the 
implementation of recommender systems that can recommend new edges.  In
the case of Slashdot, a possible application would be the recommendation
of new users to one's friend and foe list.  For these reasons, link
prediction is one of the most important machine learning task that
applies to large networks.

\subsection{Problem Formulation}
A common way to model the link prediction task is via the notion of
similarity:  If two nodes are similar, then they are likely to be
connected by an edge already, and if not, to be connected soon.  Thus, a
measure of similarity defined on a given edge set can be used to
implement a link prediction algorithm.  
We will therefore focus on the
definition of similarity measures between nodes:  functions that, given
two nodes in the network, return a similarity value.

Similarity functions are usually symmetric.  Also, they are often
positive semi-definite.  If both conditions are met, a similarity
function is a \emph{kernel}. 
However, similarity functions need not necessarily be symmetric and
positive semi-definite in our setting:  We may predict a link between
$u$ and $v$ without predicting a link between $v$ and $u$.  In this
case, we may call such a function a proximity measure.  

\subsection{Baseline Algorithms}
As baseline prediction algorithms, we use the three following
strategies: 
\begin{itemize}
\item ($1$) Always predict a positive edge.  This strategy has an
  accuracy greater than zero because there are more positive than
  negative edges.
\item ($A^T$) If there is an edge in the opposite direction, predict
  the sign of that edge.  Else predict a positive edge.  
\item ($A^2$) Use the squared adjacency matrix for prediction.  This
  simple strategy makes use of multiplicative transitivity. 
\end{itemize}
We found similarity measures based on the shortest-path distance to have no better
accuracy than these three.
The accuracy of the baseline algorithms is shown in
Table~\ref{tab:base}. 

\begin{table}[h!]
  \caption{The three baseline algorithms for link sign prediction.  The
  accuracy is measured on a scale from $-1$ to $+1$.  Greater values
  denote higher prediction accuracy.}
  \centering
  \begin{tabular}{|l||l|}
    \hline
    $1$ & 0.517 \\
    $A^T$ & 0.536 \\
    $A^2$  &  0.552 \\
    \hline
  \end{tabular}
  \label{tab:base}
\end{table}

\subsection{Algebraic Similarity Measures}
In the previous section, we used the square of the adjacency matrix $A$
to make predictions about the sign of new edges.  While calculating the
square of $A$ is possible given the corpus' size, computing the cube of $A$
is already too expensive:  While $A$ itself is sparse, $A^3$ contains
non-zero values for all node pairs that are separated by at most three
edges.  Given that the graph's radius is three and its diameter is six,
computing other powers of $A$ after $A^2$ results in non-sparse
matrices, which are too big for practical calculations.  Therefore, we
resort to dimensionality reduction.

\subsubsection{Dimensionality Reduction (A)}
The matrix $A$ can be reduced dimensionally by performing a sparse
singular value decomposition, resulting in a low-rank approximation of
the original matrix:
\begin{eqnarray}
A &\approx& A_k = U_k D_k V_k^T
\end{eqnarray}
$U_k$, $D_k$ and $V_k$ are of dimensions $n \times k$, $k\times k$ and
$n \times k$ respectively, and $D_k$ is symmetric.  The integer $k$ is the reduced dimension.

While the matrix $A$ contains zero entries for edges not present in
the original graph, the approximation $A_k$ is nonzero at these entries,
and the sign of these entries can be used as a prediction for the sign of
missing edges.
In contrast to calculating sparse matrix powers as we do for our
baseline algorithms, dimensional reduction is efficient and can
be applied to large, sparse matrices. 

\subsubsection{Symmetric Dimensionality Reduction (A sym)}
As shown in the previous section, using edges going in the opposite
direction gives more accurate prediction than only predicting positive
edges.  Therefore, we apply dimensionality reduction to the symmetric
matrix $A + A^T$.  In the case of symmetric matrices, we use the
eigenvalue decomposition:
\begin{eqnarray}
A + A^T &\approx& U_k D_k U_k^T
\end{eqnarray}
Just as for (asymmetric) dimensionality reduction, the integer $k$ represents
a parameter of the prediction algorithm.

\subsubsection{Matrix Exponential (A exp, A sym exp)}
The exponential kernel is a similarity measure based on the matrix
exponential function~\cite{b137}.  It is based on the observation that,
analogously to the real exponential, the matrix exponential can be
expressed as an infinite sum of matrix powers, with weights that decay
with the inverse factorial:
\begin{eqnarray}
\exp (A) &=& \sum_{i=0}^\infty \frac 1 {i!} A^i
\end{eqnarray}
Because the $n$-th power of the adjacency matrix of a graph contains, for
each node pair, the number of paths of length $n$ between the two nodes,
the matrix exponential represents a weighted mean of path counts between
any two nodes, with weights decaying as the inverse factorial.  
With the graph containing negative edges, the powers of $A$ represent
signed path counts, where paths with an odd number of negative edges are
counted negatively, thus implementing our \emph{enemy-of-an-enemy}
multiplication rule generalized to arbitrarily long paths. 
Although in the case of an asymmetric $A$ the exponential does not
represent a kernel, we can still use it for edge sign prediction.

Therefore, the matrix exponential will in many cases give the same
predictions for node pairs that have a distance of 2, but unlike simply
using $A^2$ will also give sensible predictions for nodes further apart,
instead of resorting to simply predicting a positive edge.  

To compute the matrix exponential, we use the fact that the matrix
exponential of a dimensionally reduced matrix can be computed by
applying the exponential to the diagonal matrix of eigenvalues.  For the
symmetric case, we have:
\begin{eqnarray}
\exp(U_k D_k U_k^T) &=& U_k \exp(D_k) U_k^T
\end{eqnarray}
The exponential of the diagonal matrix $D_k$ can be computed by applying
the real exponential to every diagonal element.  

\subsubsection{Inverted Laplacian (Ls sym)}
Laplacian kernels are defined mathematically by the pseudoinversion of
the graph's Laplacian matrix $L$.  Depending on the precise definition,
Laplacian kernels are known as resistance distance kernels~\cite{b101},
random forest kernels~\cite{b191}, random walk or mean passage time
kernels~\cite{b105} and von Neumann kernels~\cite{b263}.

Laplacian kernels have been applied to the social network analysis
problems of co-authorship graph mining~\cite{b239}, collaborative
recommender systems and community detection~\cite{b105}. 
In these cases however, the underlying graphs have only positive
edges.  In order to apply Laplacian kernels to graphs with negative
edges, we use the measure described as the signed resistance distance
in~\cite{kunegis2008a}, defined as:
\begin{eqnarray}
K &=& (\bar E - B)^+ \label{eq:lap}
\end{eqnarray}
Where $B = A + A^T$ is the symmetric adjacency matrix and
$\bar E$ is the diagonal degree matrix. 

As with the matrix exponential, the matrix pseudoinverse can be
computed easily in conjunction with dimensionality reduction, by
pseudoinverting the diagonal matrix, which amounts to inverting those diagonal
elements that are (numerically) non-zero.  

Although asymmetric (``directed'') Laplacians can be
defined~\cite{b213}, we found through extensive experimentation
that they perform poorly for our task.

\subsection{Experimental Evaluation}
To evaluate the link prediction algorithms, we split the set of edges
into a training set and a test set, train a similarity measure on the
training set, and use that similarity measure to predict the sign of the edges in the test
set.  We simply use the sign of the similarity function as the
prediction:  If the similarity function returns a positive value, we
predict a positive edge, else we predict a negative edge.  In our
evaluation, we hold out 30\% of the edges as the test set.  

We measure the
prediction accuracy on a scale from $-1$ to $+1$, where $+1$ denotes all
correct prediction and $-1$ denotes all wrong sign prediction.  Thus, an
algorithm that predicts positive and negative edges randomly with equal
probability would have an accuracy of zero. 

\begin{figure}
  \includegraphics[width=0.48\textwidth]{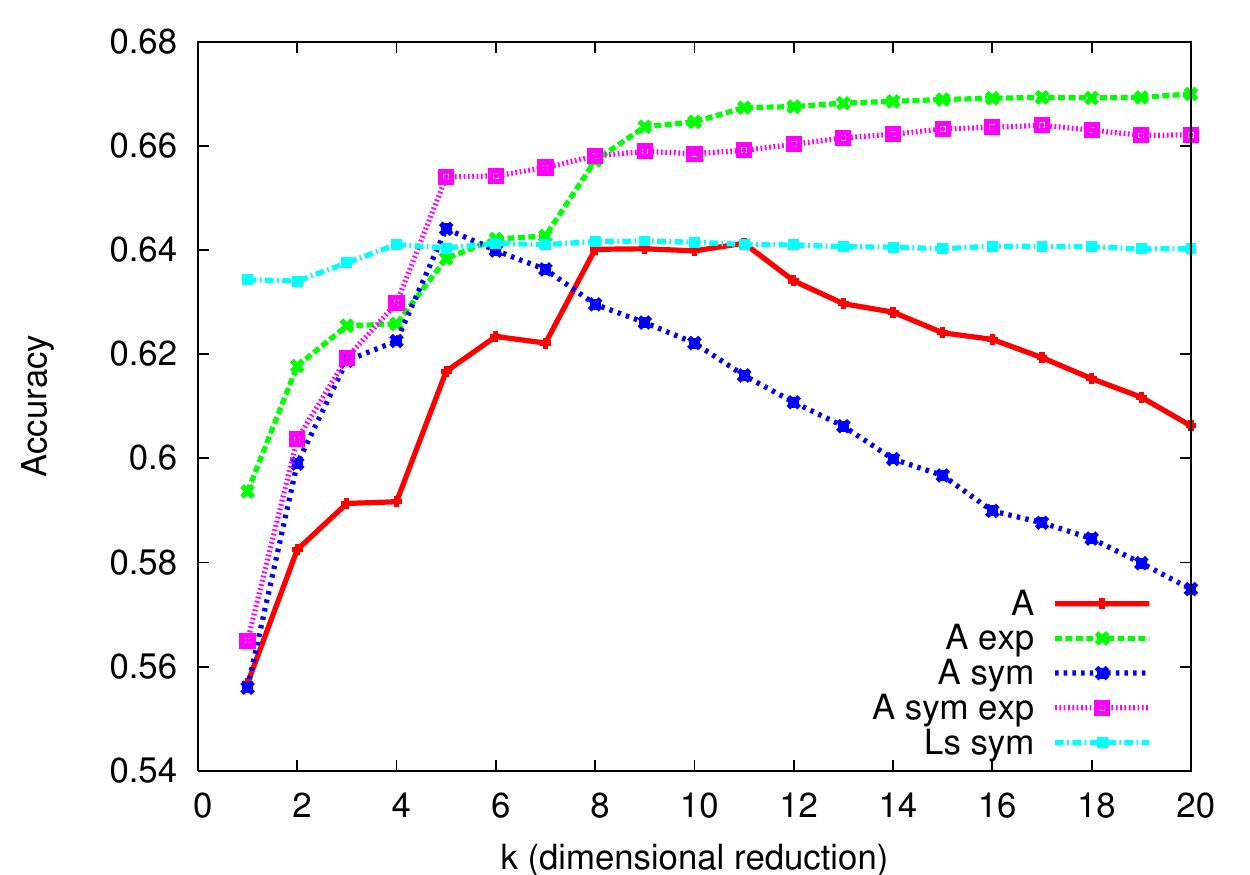}
  \caption{The accuracy of the various algebraic similarity measures on the link
  sign prediction task.  Each similarity measure is 
  tested using a varying dimensional reduction
  parameter $k$.  Greater values denote higher prediction accuracy. }
  \label{fig:pred}
\end{figure}

The evaluation results are shown in Figure~\ref{fig:pred}.  Each
similarity measure is evaluated using a varying dimensional reduction
parameter $k$.  

Our empirical study shows that the exponential kernels perform better
than their non-exponential counterparts.  Also, they do not suffer from
overfitting the data as $k$ grows:  The exponential kernels attain their
best performance asymptotically for large $k$, while \emph{A} and \emph{A sym} attain
their greatest accuracy for specific, small values of $k$.
Numerically, this is best explained by the fact that the exponential function
lets the eigenvalues of the exponential kernel become small very
fast for growing $k$.  
The overall best prediction accuracy is achieved by the asymmetric
matrix exponential. 
The good performance of the exponential kernels indicates that 
the multiplicative transitivity in the Slashdot Zoo can be extended to
paths of length greater than two if corresponding damping factors are
used, as done implicitly in the exponential kernel. 
The fact that the asymmetric exponential kernel outperforms the
symmetric exponential kernel is a hint that the assumption of symmetry
is wrong in the study of multiplicative transitivity in this signed network. 

The signed Laplacian similarity matrix provides comparable accuracy to
simple dimensionally reduction.  However, it does not suffer from
overfitting for large $k$, and it already attains it's good performance for very
small $k$. 

\section{Conclusion}
\label{sec:conclusion}
This paper considered social network analysis on graphs with negative
edge weights.  We studied
the Slashdot Zoo, a social network that is well-known for containing
negative links.
Our analysis of that social network was carried out on three levels.
On the global level, we defined the signed clustering coefficient and
relative signed clustering coefficient.
On the node level we defined Negative Rank, a new popularity measure and showed how it can be
used to identify \emph{troll} users in the Slashdot community.
On the link level, we studied the task of link sign prediction using
various signed spectral similarity measures. 
The study of the Slashdot Zoo on these three levels showed that the
network exhibits multiplicative transitivity, a property of signed
social networks that can be summarized by the phrase \emph{the enemy of my enemy is my
friend}.
We showed that these methods for analysing a network with negative edge
weights apply to large social networks and reveal facts that cannot be
uncovered using common, unsigned techniques.

Currently, we analyse further social networks with positive and
negative endorsements. 
The software running Slashdot is called
\emph{Slash}\footnote{\url{http://www.slashcode.com/}} and is nowadays
also used on other websites such as Barrapunto, Slashdot Japan, and Use
Perl.  These sites also use the Zoo feature, and could be analysed just
as the Slashdot Zoo.  
Another source of negative links in social
networks are sites such as Digg where users can rate
content by other users, giving rise to indirect negative edges.
The sites Essembly~\cite{b270} and Epinions~\cite{b325} also represent
social networks with negative edges.   

Some social network analysis methods were not covered in this paper, but
could conceivably be extended to the signed case:  betweenness centrality
measures, spectral clustering, random graph modeling of signed social
networks, stochastic approaches for prediction (which are similar to the
resistance distance approach), and others.  

\bibliographystyle{abbrv}
\bibliography{kunegis}  
\end{document}